# Vector Higgs bosons and possible suppression of flavorchanging neutral current

Xiao Yu Qian

*Department of physics , Peking university, Beijing 100871,P.R.China*

Abstract

We replace the scalar Higgs doublet with a vector Higgs boson doublet to the unified electroweak W-S model and find most of important features of W-S model are kept unchanged only the Higgs boson now become vector bosons. Lorentz invariance has been carefully discussed. The most important challenge is there will be three massless vector Higgs bosons. The remarkable effect is the possible suppression of the flavorchanging neutral current compare to the multi-Higgs model.

PACS number(s):12.60.Fr, 12.15.Mm

Keywords:Vector Higgs bosons,Lorentz invariance,flavorchanging neutral current.

There are at least two basic problems remain open in the W-S model. The first is the Higgs boson has not been found yet. Perhaps it does not exist at all. The second is the renormalization of the flavorchanging neutral current are not very clear. Also when people add more scalar Higgs doublets to the model the calculated rate of rare decays are always in excess of the experimental bounds [1]. Now we are looking for a new solution. We try to keep all these results in W-S model which have been verified by experiments. The approach is to replace the traditional scalar Higgs doublet with a vector Higgs doublet.

We begin with the discussion of the possibility of vector Higgs bosons. We will not generally discuss this problem, instead we bound ourselves to the SU(2)×U(1) gauge model. Suppose

$$U_\mu = \begin{bmatrix} U_\mu^+ \\ U_\mu^0 \end{bmatrix} \quad (1)$$

is a SU(2)×U(1) doublet.

$$<0|U_\mu|0> = \begin{bmatrix} 0 \\ u_\mu \end{bmatrix} \quad (2)$$

In which $u_\mu$ is a c-number Lorentz 4-vector. Now the problem is whether it will spoil the Lorentz invariance as people used to think. When we go from one reference frame to another $U_\mu$ will change to $a_\mu^\nu U_\nu$ and $u_\mu$ will change to $a_\mu^\nu u_\nu$ and we easy to see the vacuum state will not change. But if we define the Lorentz transformation as a unitary transformation to the $U_\mu$

$$a_\mu^\nu U_\nu = L U_\mu L^+$$

then

$$<0|U_\mu|0> = <0|L^+LU_\mu L^+L|0>$$
$$= a_\mu^\nu <0|U_\nu|0>$$
$$= a_\mu^\nu \begin{bmatrix} 0 \\ u_\nu \end{bmatrix}$$
$$\neq \begin{bmatrix} 0 \\ u_\mu \end{bmatrix}$$

it is a contradiction. But it is incorrect. Since $U_\mu$ is not a true quantum operator instead

$$U'_\mu = U_\mu - <0|U_\mu|0> \tag{3}$$

is the true quantum fluctuation. So we should define the Lorentz transformation as a unitary operation to $U'_\mu$

$$a_\mu^\nu U'_\nu = LU'_\mu L^+ \tag{4}$$

and easy to see the vacuum is Lorentz invariant. In the following we also see the whole formalism is Lorentz invariant, then the Lorentz invariance for the whole theory is kept no problem.

People may argues that the vacuum states are infinitely degenerate and the VEV are related by Lorentz transformations continuously. Indeed the vacuum states are infinitely degenerate, but if you chose a reference frame and have the relation (2) we have shown the vacuum state is Lorentz invariant already. The vacuum state is not symmetric for the Lorentz transformation, but once it has been chosen it is Lorentz invariant. We have to discriminate two distinct problems. The first is when you choose a specific Lorentz reference frame then different $u_\mu$'s would result in different vacuum state, they are infinitely degenerate. The second is when you choose two different Lorentz reference frame, there will be two $U_\mu$'s related each other by a Lorentz transformation and you choose two VEV's also related each other by the same Lorentz transformation then the vacuum states are the same, they are not dependent on the specific choice of the reference frame. This is the key point on the problem. Then the left problem is whether we need to specify a specific value of $u_\mu$ for any specific reference frame. In most cases it is not necessary because only $u^2 = u^\mu u_\mu$ appears, no single $u_\mu$ appears. The only exception is the $U_\mu$ production, a $u_\mu$ would be singled out coupled to the final state. It is a regret point of this theory, but it does not mean Lorentz invariance is lost.

The kinetic part of the Lagrangian of $U_\mu$ is

$$\mathscr{L}_U = -(D_\mu U_\nu)^+ (D^\mu U^\nu) - V(U_\nu^+ U^\nu) \tag{5}$$

$$D_\mu U_\nu = (\partial_\mu - \frac{1}{2}ig\vec{W}_\mu \cdot \vec{\sigma} - \frac{1}{2}ig'B_\mu)U_\nu$$

$$V(U_\nu^+ U^\nu) = m^2 U_\nu^+ U^\nu + \frac{\Lambda}{4}(U_\nu^+ U^\nu)^2$$

$V(U_\nu^+ U^\nu)$ has a minimum at $U_\nu^+ U^\nu = u_\nu u^\nu = -\frac{2m^2}{\Lambda}$, where $u_\nu$ is a c-number spacelike Lorentz 4-vector. $\mathscr{L}_U$ has a contribution to $\vec{W}_\mu$ and $B_\mu$ mass term

$$-\frac{g^2}{4}u_\nu u^\nu(W_1^2 + W_2^2) - \frac{g^2}{4}u_\nu u^\nu W_3^2 - \frac{g'^2}{4}u_\nu u^\nu B^2 + \frac{1}{2}gg'u_\nu u^\nu W_3 B$$

it's similar to the W-S model case, the only difference is that $-u_\nu u^\nu$ replaced $\frac{v^2}{2}$. The relation $M_Z = \frac{M_W}{\cos\theta_W}$ is kept unchanged and $-\frac{1}{2}g^2 u_\nu u^\nu = M_W^2$. After spontaneous symmetry breaking expand $V(U_\nu^+ U^\nu)$ to power 2 terms of the field quantities

$$V = \frac{\Lambda}{2}(u_\nu U_1^{0\nu})^2 + \cdots \tag{6}$$

where $U_{1\nu}^0 = \frac{1}{\sqrt{2}}(U_\nu^{'0*} + U_\nu^{'0})$. So three of the $U_\mu$ become massless and lost three degrees of freedom. The three degrees of freedom were absorbed by $W_\mu^\pm$ and $Z_\mu$. The three massless vector bosons is a challenge to updated experimental data. Also it is the most important test to this theory. On the other hand (6) is not a regular mass term for $U_{1\mu}^0$, but we can look (6) as a two points interaction

$$\mathscr{L}_I = -\frac{\Lambda}{2}(u_\nu U_1^{0\nu})^2$$

Then generate mass of $U_{1\mu}^0$ by renormalization. We begin with a massless propagator $\frac{-ig_{\mu\nu}}{p^2}$ for $U_{1\mu}^0$, then calculated the complete propagator of $U_{1\mu}^0$, the result is

$$\frac{-i[(1+\frac{\Lambda u^2}{p^2})g_{\mu\nu} - \frac{\Lambda u_\mu u_\nu}{p^2}]}{p^2 + \Lambda u^2} \tag{7}$$

There seems have two poles 0 and $-\Lambda u^2$, but if multiplies it by $u^\mu$ it becomes

$$\frac{-iu_\nu}{p^2 + \Lambda u^2}$$

So the pole 0 can be eliminated, the only pole is $-\Lambda u^2 \equiv -\Lambda u_\nu u^\nu$, it is the mass of $U^0_{1\mu}$. It is similar with the traditional W-S model that the mass of vector Higgs particles $U^0_{1\mu}$ is undetermined.

The Lagrangian of $U_\mu$-quark coupling can been written as

$$\mathscr{L}_{UQ} = g_{ij}\overline{\Psi}_{Li}u_{Rj}u_\mu\tilde{U}^\mu + g_{ij}\overline{u}_{Ri}u_\mu\tilde{U}^{\mu+}\Psi_{Lj} + g'_i\overline{\Psi}_{Li}d_{Ri}u_\mu U^\mu + g'_i\overline{d}_{Ri}u_\mu U^{\mu+}\Psi_{Li} \quad (8)$$

There $u_\mu$ is a c-number 4-vector just equal to the VEV of $U_\mu$, exactly speaking it is a SU(2)×U(1) singlet. It just plays a role like a coupling constant. After spontaneous symmetry breaking it gives a mass term to 6 quarks, it is

$$\overline{u}_i g_{ij} u_j u^2 + g'_i \overline{d}_i d_i u^2 = -(\overline{u}\ \overline{c}\ \overline{t})\begin{pmatrix} m_u & & \\ & m_c & \\ & & m_t \end{pmatrix}\begin{pmatrix} u \\ c \\ t \end{pmatrix} - (\overline{d}\ \overline{s}\ \overline{b})\begin{pmatrix} m_d & & \\ & m_s & \\ & & m_b \end{pmatrix}\begin{pmatrix} d \\ s \\ b \end{pmatrix}$$

where $\begin{pmatrix} u \\ c \\ t \end{pmatrix} = V\begin{pmatrix} u_1 \\ u_2 \\ u_3 \end{pmatrix}$, $V$ is the CKM matrix.

For the $U^\pm_\mu$-quark coupling we have

$$-g_{ij}\overline{d}_{Li}u_{Rj}u_\mu U^{-\mu} - g_{ij}\overline{u}_{Ri}d_{Lj}u_\mu U^{+\mu} + g'_i\overline{u}_{Li}d_{Ri}u_\mu U^{+\mu} + g'_i\overline{d}_{Ri}u_{Li}u_\mu U^{-\mu}$$

$$= (\overline{d}_L\ \overline{s}_L\ \overline{b}_L)V^+\begin{pmatrix} m_u & & \\ & m_c & \\ & & m_t \end{pmatrix}\begin{pmatrix} u_R \\ c_R \\ t_R \end{pmatrix}\frac{u_\mu U^{-\mu}}{u^2}$$

$$+ (\overline{u}_R\ \overline{c}_R\ \overline{t}_R)\begin{pmatrix} m_u & & \\ & m_c & \\ & & m_t \end{pmatrix}V\begin{pmatrix} d_L \\ s_L \\ b_L \end{pmatrix}\frac{u_\mu U^{+\mu}}{u^2}$$

$$- (\overline{u}_L\ \overline{c}_L\ \overline{t}_L)V\begin{pmatrix} m_d & & \\ & m_s & \\ & & m_b \end{pmatrix}\begin{pmatrix} d_R \\ s_R \\ b_R \end{pmatrix}\frac{u_\mu U^{+\mu}}{u^2}$$

$$- (\overline{d}_R\ \overline{s}_R\ \overline{b}_R)\begin{pmatrix} m_d & & \\ & m_s & \\ & & m_b \end{pmatrix}V^+\begin{pmatrix} u_L \\ c_L \\ t_L \end{pmatrix}\frac{u_\mu U^{-\mu}}{u^2}$$

$$= -\begin{pmatrix} \bar{d}_L & \bar{s}_L & \bar{b}_L \end{pmatrix} V^+ \begin{pmatrix} m_u u_R \\ m_c c_R \\ m_t t_R \end{pmatrix} \frac{g^2 u_\mu U^{-\mu}}{2M_W^2} - \begin{pmatrix} m_u \bar{u}_R & m_c \bar{c}_R & m_t \bar{t}_R \end{pmatrix} V \begin{pmatrix} d_L \\ s_L \\ b_L \end{pmatrix} \frac{g^2 u_\mu U^{+\mu}}{2M_W^2}$$

$$+ \begin{pmatrix} \bar{u}_L & \bar{c}_L & \bar{t}_L \end{pmatrix} V \begin{pmatrix} m_d d_R \\ m_s s_R \\ m_b b_R \end{pmatrix} \frac{g^2 u_\mu U^{+\mu}}{2M_W^2} + \begin{pmatrix} m_d \bar{d}_R & m_s \bar{s}_R & m_b \bar{b}_R \end{pmatrix} V^+ \begin{pmatrix} u_L \\ c_L \\ t_L \end{pmatrix} \frac{g^2 u_\mu U^{-\mu}}{2M_W^2}$$

(9)

For the two neutral vector Higgs bosons and quarks coupling we have

$$g_{ij} \bar{u}_{Li} u_{Rj} u_\mu U'^{0*\mu} + g_{ij} \bar{u}_{Ri} u_{Lj} u_\mu U'^{0\mu} + g'_i \bar{d}_{Li} d_{Ri} u_\mu U'^{0\mu} + g'_i \bar{d}_{Ri} d_{Li} u_\mu U'^{0*\mu}$$

$$= -\begin{pmatrix} \bar{u} & \bar{c} & \bar{t} \end{pmatrix} \begin{pmatrix} m_u & & \\ & m_c & \\ & & m_t \end{pmatrix} \begin{pmatrix} u \\ c \\ t \end{pmatrix} \frac{u_\mu U_1^{0\mu}}{\sqrt{2} u^2}$$

$$- \begin{pmatrix} \bar{d} & \bar{s} & \bar{b} \end{pmatrix} \begin{pmatrix} m_d & & \\ & m_s & \\ & & m_b \end{pmatrix} \begin{pmatrix} d \\ s \\ b \end{pmatrix} \frac{u_\mu U_1^{0\mu}}{\sqrt{2} u^2}$$

$$+ i \begin{pmatrix} \bar{u} & \bar{c} & \bar{t} \end{pmatrix} \gamma^5 \begin{pmatrix} m_u & & \\ & m_c & \\ & & m_t \end{pmatrix} \begin{pmatrix} u \\ c \\ t \end{pmatrix} \frac{u_\mu U_2^{0\mu}}{\sqrt{2} u^2}$$

$$- i \begin{pmatrix} \bar{d} & \bar{s} & \bar{b} \end{pmatrix} \gamma^5 \begin{pmatrix} m_d & & \\ & m_s & \\ & & m_b \end{pmatrix} \begin{pmatrix} d \\ s \\ b \end{pmatrix} \frac{u_\mu U_2^{0\mu}}{\sqrt{2} u^2}$$

$$= (m_u \bar{u} u + m_c \bar{c} c + m_t \bar{t} t) \frac{g^2 u_\mu U_1^{0\mu}}{2\sqrt{2} M_W^2} - i(m_u \bar{u} \gamma^5 u + m_c \bar{c} \gamma^5 c + m_t \bar{t} \gamma^5 t) \frac{g^2 u_\mu U_2^{0\mu}}{2\sqrt{2} M_W^2}$$

$$+ (m_d \bar{d} d + m_s \bar{s} s + m_b \bar{b} b) \frac{g^2 u_\mu U_1^{0\mu}}{2\sqrt{2} M_W^2} + i(m_d \bar{d} \gamma^5 d + m_s \bar{s} \gamma^5 s + m_b \bar{b} \gamma^5 b) \frac{g^2 u_\mu U_2^{0\mu}}{2\sqrt{2} M_W^2}$$

(10)

Exchange a $U_{1\mu}^0$ or $U_{2\mu}^0$ will not result in any change in flavor. But exchange a $W_\mu^\pm$ or $U_\mu^\pm$ will create a propagator such as from d-quark to s-quark and the vertex such as from d-quark and anti-s-quark to $Z^0$ or $\gamma$. There has nobody have made any precise theory to such renormalization. So we cannot make any prediction to the renormalized flavorchanging neutral current coupling constant too. But compare to the multi scalar Higgs model, exchange a $U_\mu^\pm$ will introduce a negative sigh to the loop integral from the factor $u^\mu u_\mu$, and the orders of magnitude of the integrals are almost same. Reasonably we can expect the flavorchanging neutral current will be effectively suppressed.

Since the emerging of W-S model it's almost 40 years passed but the Higgs boson still have not been found. This model maybe shed a drop of new light to the sensitive physics area.

Finally I have to say some place in this paper mathematically it is not rigorous. The quantization of massless vector boson is a problem remain open. I would like to leave this problem to mathematicians. But the idea of this paper is interesting and important.